\documentclass[letterpaper]{article} % DO NOT CHANGE THIS
\usepackage{aaai24}  % DO NOT CHANGE THIS
\usepackage{times}  % DO NOT CHANGE THIS
\usepackage{helvet}  % DO NOT CHANGE THIS
\usepackage{courier}  % DO NOT CHANGE THIS
\usepackage[hyphens]{url}  % DO NOT CHANGE THIS
\usepackage{graphicx} % DO NOT CHANGE THIS
\usepackage{natbib}  % DO NOT CHANGE THIS AND DO NOT ADD ANY OPTIONS TO IT
\usepackage{caption} % DO NOT CHANGE THIS AND DO NOT ADD ANY OPTIONS TO IT
\usepackage{algorithm}
\usepackage{algorithmic}

\usepackage{caption,subcaption}
\usepackage{amsmath, amssymb, amsfonts}
\usepackage{xcolor}
\usepackage{booktabs, makecell, multirow, tabularx}
\usepackage{colortbl}
\usepackage{url}
\usepackage{array}
\usepackage{newfloat}
\usepackage{listings}
\DeclareCaptionStyle{ruled}{labelfont=normalfont,labelsep=colon,strut=off} % DO NOT CHANGE THIS
\floatstyle{ruled}
\newfloat{listing}{tb}{lst}{}
\floatname{listing}{Listing}
\title{Learning Time Slot Preferences via Mobility Tree for Next POI Recommendation}
\author{
    Tianhao Huang\textsuperscript{\rm 1}\equalcontrib,
    Xuan Pan\textsuperscript{\rm 1}\textsuperscript{\rm 2}\equalcontrib,
    Xiangrui Cai\textsuperscript{\rm 1}\textsuperscript{\rm 3}\textsuperscript{\rm 4}\thanks{Corresponding author.},
    Ying Zhang\textsuperscript{\rm 1}\textsuperscript{\rm 3},
    Xiaojie Yuan\textsuperscript{\rm 1}\textsuperscript{\rm 2}\textsuperscript{\rm 3}
}
\affiliations{
    \textsuperscript{\rm 1}College of Computer Science, Nankai University \\
    \textsuperscript{\rm 2}Tianjin Key Laboratory of Network and Data Security Technology, Tianjin, China \\
    \textsuperscript{\rm 3}Key Laboratory of Data and Intelligent System Security, Ministry of Education, China \\
    \textsuperscript{\rm 4}Science and Technology on Communication Networks Laboratory, Shijiazhuang, China\\
    tianhao.alex.huang@gmail.com,
    panxuan@dbis.nankai.edu.cn, 
    \{caixr, yingzhang, yuanxj\}@nankai.edu.cn
}

\usepackage{bibentry}
\begin{document}

\maketitle

\begin{abstract}
Next Point-of-Interests (POIs) recommendation task aims to provide a dynamic ranking of POIs based on users' current check-in trajectories. The recommendation performance of this task is contingent upon a comprehensive understanding of users' personalized behavioral patterns through Location-based Social Networks (LBSNs) data. While prior studies have adeptly captured sequential patterns and transitional relationships within users' check-in trajectories, a noticeable gap persists in devising a mechanism for discerning specialized behavioral patterns during distinct time slots, such as noon, afternoon, or evening. In this paper, we introduce an innovative data structure termed the ``Mobility Tree'', tailored for hierarchically describing users' check-in records. The Mobility Tree encompasses multi-granularity time slot nodes to learn user preferences across varying temporal periods. Meanwhile, we propose the Mobility Tree Network (MTNet), a multitask framework for personalized preference learning based on Mobility Trees. We develop a four-step node interaction operation to propagate feature information from the leaf nodes to the root node. Additionally, we adopt a multitask training strategy to push the model towards learning a robust representation. The comprehensive experimental results demonstrate the superiority of MTNet over ten state-of-the-art next POI recommendation models across three real-world LBSN datasets, substantiating the efficacy of time slot preference learning facilitated by Mobility Tree.
\end{abstract}

\section{Introduction}
The advent of location-based social networks (LBSNs) facilitate users to share their geographical locations.
The huge amount of geographical data opens up new opportunities to learn user preferences and recommend Point-of-Interests (POIs) to users. Next POI recommendation task aims to provide a ranked list of POIs that users are most likely to visit in the future, by discerning the dynamic preferences of users \cite{manotumruksa2018contextual} through their current check-in trajectories.
It is beneficial for both user time scheduling and business expansion \cite{liu2017experimental}.

\begin{figure}[t]
    \centering
    \includegraphics[width=0.98\columnwidth]{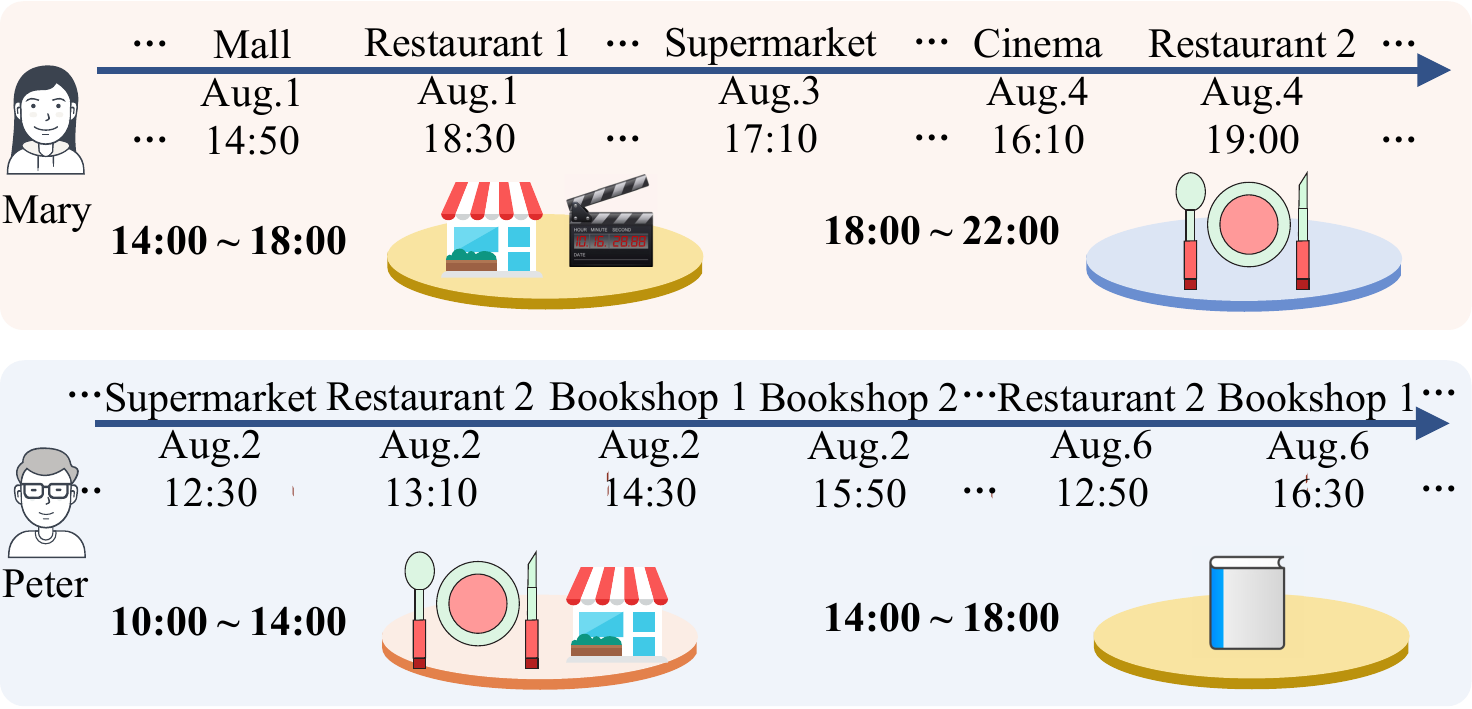}
    \caption{Mary and Peter exhibit varying preferences that evolve over distinct periods in one day. For example, during the period of ``14:00$\sim$18:00'', Mary prefers to go shopping in a mall or watch a movie in a cinema, whereas Peter likes to visit several bookstores.}
    \label{intro}
\end{figure}

Previous studies on next POI recommendation organize a trajectory of a user by a sequence of check-ins, where the items are arranged in chronological order.
%Most next POI recommendation models employ the organization of users' check-in data in chronological order, thereby delineating check-in sequences.
%
This approach promotes the comprehension of sequential patterns and transitional dependencies, enabling the capture of users' short-term preferences \cite{feng2018deepmove, manotumruksa2018contextual, huang2019attention, guo2020attentional}.
Some attempts such as LSTPM \cite{sun2020go} and STGN \cite{7} have boosted recommendation performance by integrating long-term and short-term user preference modeling strategies.
To further enhance check-in communication, STAN \cite{9} adopted the self-attention mechanism to facilitate interactions among non-consecutive check-ins.
Recently, more and more studies \cite{rao2022graph, 10, lim2022hierarchical, wang2021attentive} have leveraged graph structure in check-in descriptions to exploit %develop 
collaborative signals among different users and capture global transition relationships.

While prior research has made considerable advancements in modeling sequential patterns and transitional relationships within trajectories, they primarily derive user preferences based solely on the sequencing of check-ins, thereby ignoring the personalized preferences across discrete time slots.
We observe that users frequent specific POIs during a relatively fixed period, rather than a specific time point.
%Upon observing the check-in records, we find that the user consistently frequents specific POIs during relatively fixed time intervals.
%
%This observation is exemplified by 
As illustrated in \figurename~\ref{intro}, wherein Mary's propensity to visit a mall or a supermarket manifests not as a rigid adherence to a precise time point but rather within the broader temporal window from 14:00 to 18:00.
However, Peter's inclination toward several bookshops is observed within the same 14:00 to 18:00 window.
Notably, Mary's preference for shopping at malls does not appear as pronounced during other temporal periods.
This lends credence to the assumption that individualized preferences can be delineated even further at the temporal granularity level.

Moreover, user behavior patterns are not uniformly consistent throughout each day.
For example, Mary's trajectory on a leisure day encompassing ``Cinema -- Mall -- Supermarket -- Restaurant'', which contrasts with her routine on a working day, organized by ``Company -- Café -- Supermarket -- Apartment''. In the event that the last check-in within a given trajectory is visiting a supermarket, it is observed that divergent next POIs emerge across distinct days.
On a rest day, it is a ``Restaurant'', but on a workday, it is the ``Apartment''.
To summarize, the temporal variability underscores the necessity of subdividing user preferences beyond daily granularity and extending into specific temporal segments.
Regrettably, existing methodologies lack any endeavors toward segregating user preferences across divergent time slots with varying granularities when modeling the check-in trajectories.

In this paper, we propose a hierarchical structure, referred to as the ``Mobility Tree'', as an innovative approach to encapsulate users' check-in records.
Different from the prevalent sequential-based or graph-based structures, the Mobility Trees are constructed by integrating multi-granularity time slot nodes.
Each time slot node is tailored to aggregate check-in occurrences within a specific time period.
This distinctive construction enables the discrimination of user preferences across different temporal phases,
% by dedicated time bucket nodes, 
thus enhancing the holistic understanding of users' varied behavioral patterns.
In alignment with the formulation of the Mobility Tree concept, we accordingly introduce the Mobility Tree Network (MTNet) to learn users' dynamic preferences for the next POI recommendation task.
To adapt the specific topological structure of the Mobility Tree, we devise a four-step node interaction operation in MTNet for facilitating information propagation from raw check-in records towards the hierarchical time slot nodes.
We adopt a multitask training strategy to enhance the representation ability and robustness of the Mobility Trees.
This strategy orchestrates a collaborative prediction for the next POI and the corresponding contextual information.
The weights of multitasks are adjusted by a self-adaptive approach.

In summary, the main contributions of this paper are as follows:
\begin{itemize}
  \item We introduce a novel hierarchical check-in description method named Mobility Tree. Concretely, the trees consist of multi-granularity time slot nodes to capture users' distinct preferences across diverse periods that have been ignored in previous works.
 \item We propose Mobility Tree Network (MTNet) to grasp users' dynamic preferences in the next POI recommendation task. In particular, we devise a four-step node interaction operation for message passing in Mobility Trees and adopt a multitask training strategy to push towards learning a robust representation.
 \item Extensive experiments are conducted on three real-world LBSN datasets. The results demonstrate the superiority of MTNet when compared to ten state-of-the-art baselines. We also provide in-depth analysis of the proposed model via ablation study and visualization analysis.
\end{itemize}

\section{Related Work}
POI Recommendation is a popular service in LBSNs, and the next POI recommendation is a typical and well-studied branch \cite{sanchez2022point}.
The fundamental assumption in this problem is that users' future movements and activities are strongly influenced by their latest check-in behaviors \cite{zhang2022next}.

Currently, the prevailing paradigm for capturing user preferences involves encoding user check-ins based on sequential-based models, such as Recurrent Neural Networks (RNNs).
Several extensions of the traditional RNN model have been proposed to enhance the representation of check-in features and user preference comprehension.
For example, ST-RNN \cite{6} incorporates spatial-temporal transition matrices within the recurrent structure to capture the check-in features.
CARA \cite{manotumruksa2018contextual} improves upon gating mechanisms that control the influence of contextual information on hidden states between recurrent units.
LSTPM \cite{sun2020go} and STGN \cite{7} focus on modeling long-term and short-term preferences within LSTM-based architectures to enhance the correlation between check-ins.

Recently, attention mechanisms have been employed in various sequential-based models to capture high-order dependencies among check-ins.
For instance, ARNN \cite{guo2020attentional} integrates RNNs with attention layers to select highly salient neighbors that are correlated with the current check-in at each time step.
ATST-LSTM \cite{huang2019attention} introduces an attention-based spatiotemporal LSTM network that uses contextual information to highlight relevant historical check-ins in a sequence.
In recent years, Transformers \cite{vaswani2017attention} have gained popularity across multiple domains due to their effectiveness in modeling long-range dependencies.
GETNext \cite{10} adopts a transformer framework to integrate multiple elements into the preference representation.
GeoSAN \cite{lian2020geography} proposes a self-attention-based geography encoder to capture spatial proximity between nearby locations.

Some studies construct graphs to exploit global transition patterns for check-in interactions across different trajectories, such as GETNext \cite{10}, HMT-GRN \cite{lim2022hierarchical}, ASGNN \cite{wang2021attentive}, and Graph-Flashback \cite{rao2022graph}.
However, in the aforementioned models, the check-in context, such as geography and category information, is only considered as POI features and not decoupled as part of the user's mobility behavior within the check-in sequence.
In this paper, we propose to encode check-in context based on tree structures instead of sequences and graphs to provide a more detailed description of user mobility behavior in the real world.

\section{Problem Formulation}
In this section, we introduce the related definition and the problem formulation of next POI recommendation.

We denote the set of users as $\mathcal{U}=\left\{u_{1}, u_{2}, \ldots, u_{|\mathcal{U}|}\right\}$, the set of POIs as $\mathcal{L}=\left\{l_{1}, l_{2}, \ldots, l_{|\mathcal{L}|}\right\}$, and the set of timestamps as $\mathcal{T}=\left\{t_{1}, t_{2}, \ldots, t_{|\mathcal{T}|}\right\}$. Additionally, we define the list of POI categories (e.g., cafe or restaurant) as set $\mathcal{C}=\left\{c_{1}, c_{2}, \ldots, c_{|\mathcal{C}|}\right\}$. Here, each POI $l_{i} \in \mathcal{L}$ encompasses both geographic information (latitude and longitude of the POI) and categorical information, which is represented by a tuple $\langle\text {lat, lon, cat}\rangle$. On this basis, we put forth several definitions for the problem of next POI recommendation.

\textbf{DEFINITION 1: (Check-in)} A check-in is denoted as a tuple $s = \langle {u, l, t}\rangle \in \mathcal{U} \times \mathcal{L} \times \mathcal{T}$, which indicates that user $u$ visited venue $l$ at timestamp $t$.

% \textbf{DEFINITION 2: (Trajectory)} A trajectory of user $u\in \mathcal{U}$ is denoted by a sequence $S^u = \left\{s^u_1, s^u_2, \dots, s^u_t\right\}$, where $|s^u_t|$ refers to the length of the check-in sequences.
% We denote the trajectory set of all the users by $\mathcal S=\left\{ S^{u_1}, S^{u_2}, \dots, S^{u_{|\mathcal{U}|}} \right\}$.
% Note that we split all check-ins of a user to several trajectories, which is similar to previous studies \cite{sun2020go, 10}.

\textbf{DEFINITION 2: (Trajectory)} For each user $u\in \mathcal{U}$, we split all check-ins of the user $u$ to a trajectory sequence denoted by $\mathcal{S}^{u}=\left\{ S^{u}_{1}, S^{u}_{2}, \dots, S^{u}_{n} \right\}$. Each trajectory $S^{u}_{m}\in \mathcal{S}^{u}$ comprises a sequence of check-ins visited by the user $u$ in a chronological order, i.e., $S^{u}_{m} = \left\{s^u_1, s^u_2, \dots, s^u_{k}\right\}$ and $k$ is the index of the last check-in of trajectory $S^{u}_{m}$.

\textbf{DEFINITION 3: (Next POI recommendation)} 
Given the current trajectory of user $u$, i.e. $S^{u}_{m} = \left\{s^u_1, s^u_2, \dots, s^u_{k}\right\}$, the objective of next POI recommendation is to recommend top-$k$ POIs that the user is most likely to visit in the future.
% at the $t+1$ timestamps.
Specifically, the recommendation model generates a list of probabilities for all candidates $\mathcal{P}=\left\{p_{1}, p_{2}, \ldots, p_{|\mathcal{L}|}\right\}$, and then return the top-$k$ POIs with the highest probabilities from the list $\mathcal{P}$ for recommendation.

\section{Method}
In this section, we introduce the proposed next POI recommendation model, MTNet. It consists of four modules corresponding to the Mobility Tree construction, node initialization, node information interaction, and multitask learning.

\subsection{Mobility Tree Construction}
\begin{figure}[t]
    \centering
    \includegraphics[width=0.8\columnwidth]{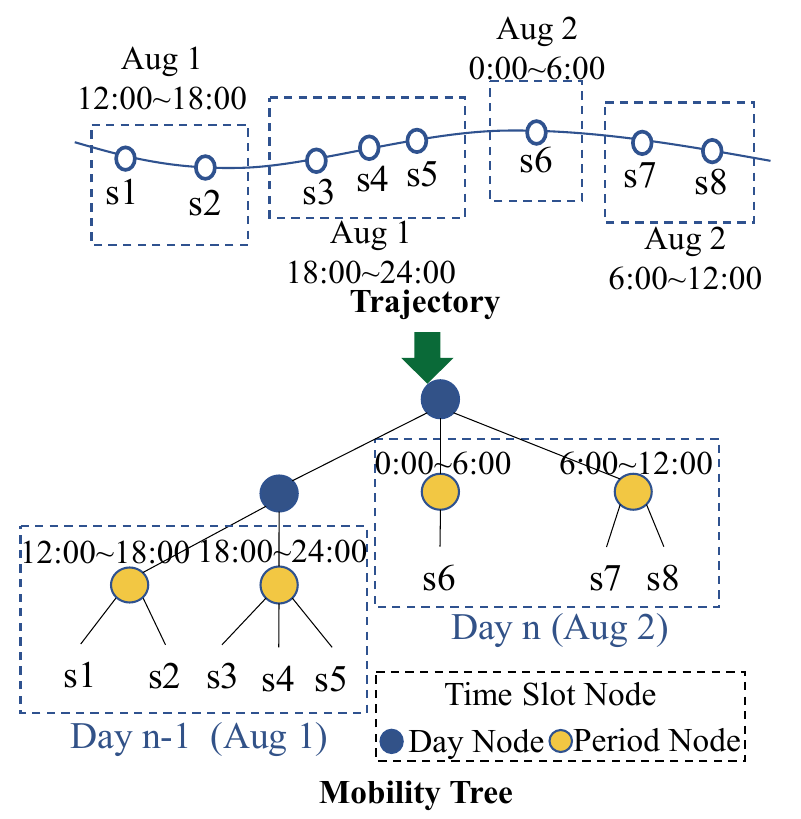}
    \caption{Illustration of a Mobility Tree construction.}
    \label{fig_tree_construction}
\end{figure}

As discussed in the introduction, the purpose of building Mobility Trees is to especially represent and learn users' personalized preferences hidden in varied time slots, including a whole day and specific periods each day. As shown in \figurename~\ref{fig_tree_construction}, given a trajectory of user $u$, i.e., $S^{u}_{m} = \left\{s^u_1, s^u_2, \dots, s^u_{k}\right\}$, we can illustrate the corresponding Mobility Tree $TR$. It consists of the nodes representing the multi-granularity time slots and the raw check-ins. More specifically, the blue nodes are the coarse-grained time slot nodes, which integrate the trajectory of a user in one day, called the day nodes.
The child nodes of the day nodes, that is, the yellow ones, are the fine-grained time slot nodes, describing the trajectory of a user in a certain period of time, called the period nodes.
The child nodes of the period nodes are the raw check-in nodes, representing the check-in information from the LBSN data.
All the raw check-in nodes are the leaves of the Mobility Trees since they are no longer subdivided.

Taking the trajectory in \figurename~\ref{fig_tree_construction} as an example, the user has a trajectory across two days.
$\left\{s_{1}, s_{2}, \ldots, s_{5}\right\}$ are the check-ins from Aug. 1, and $\left\{s_{6}, s_{7}, s_{8}\right\}$ are the check-ins from Aug. 2.
If we divide the day into four periods, each day has the periods of ``0:00~6:00,'' ``6:00~12:00,'' ``12:00~18:00,'' and ``18:00~24:00''.
Therefore we can divide the check-ins from the two days into four time periods.
Then we can construct the Mobility Tree according to the arrangement of the time periods as the time slot nodes and the raw check-in nodes, as shown in the figure.

\subsection{Node Initialization} \label{sec:nim}
We employ the embedding layers to encode the POI, user, geographical location, category, and time slot into latent representations as $\mathbf{e}^l \in \mathbb{R}^d, \mathbf{e}^u \in \mathbb{R}^d, \mathbf{e}^g \in \mathbb{R}^d, e^c \in \mathbb{R}^d, \mathbf{e}^t \in \mathbb{R}^{4d}$, respectively. Here, for the timestamp of each check-in record, we divide one day into 24 time slots, corresponding to 24 hours, helping better represent the periodicity of check-in records. Moreover, the location information of the POIs is converted to the areas clustered by the k-means method. We denote $|\mathcal{G}|$ as the cluster number. Then, we assign areas with unique IDs and allocate the IDs to all the POIs' locations based on the area they belong to.

For each trajectory $S$, we initialize the raw check-in nodes as the concatenation of the embeddings, then we attach the time slot to the $e_{s}$ as follows:
% the output of the multi-model embedding layer is the concatenation of all the embeddings above, which is denoted as:
\begin{equation}
    \mathbf{e}_{s} = [\mathbf{e}^u;\mathbf{e}^l;\mathbf{e}^c;\mathbf{e}^g] + \gamma \mathbf{e}^t\text {, }
\end{equation}
where $[;]$ denotes the concatenation operation, $\gamma$ is for controlling the influence of the 
visiting time, and $e_{s}$ id the embedding of the raw check-in node of $s \in S$.
% Then we attach the time slot to the $e_{s}$ as follows:
% \begin{equation}
%     e_{s}^{\left(0\right)} = e_{s} + \alpha e^t,
% \end{equation}

\subsection{Node Information Interaction}

\begin{figure}[t]
    \centering
    \includegraphics[width=0.9\columnwidth]{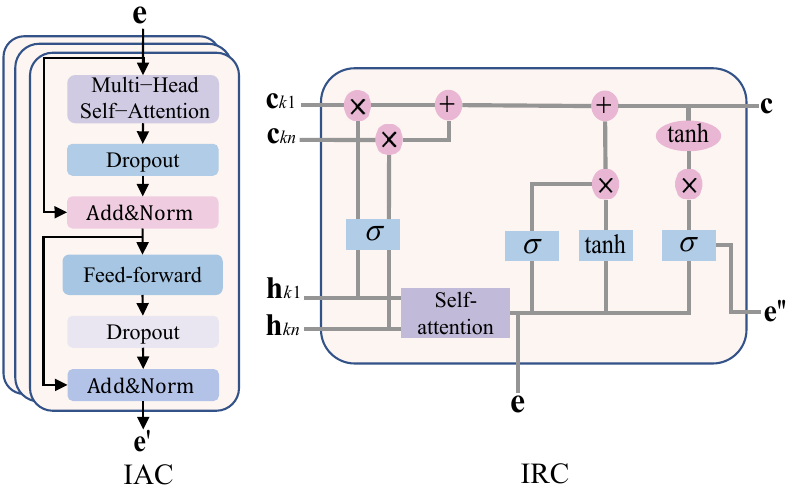}
    \caption{Network structures of Intra-hierarchy Communication (IAC) and Inter-hierarchy Communication (IRC).}
    \label{fig_IAC_IRC}
\end{figure}

Before providing the two-stage message-passing mechanism, we first introduce two basic operations, whose network structures are shown in \figurename~\ref{fig_IAC_IRC}.
The first one is Intra-hierarchy Communication, IAC for short.
It is used to realize the information exchange among the child nodes belonging to the same parent node.
Given the child nodes $\mathbf{E}$, we adopt a self-attention encoder \cite{vaswani2017attention} that includes multi-head self-attention layers to transform each $\mathbf{e} \in \mathbf{E}$ as follows:
\begin{equation}
\begin{aligned}
\mathbf{e}_{i j}^{(h)} & =\frac{\mathbf{e}_i \mathbf{W}_Q^{(h)}\left(\mathbf{e}_j \mathbf{W}_K^{(h)}\right)^{\top}}{\sqrt{d_z / H}} \text {, } \; \alpha_{i j}^{(h)}=\frac{\exp \left(\mathbf{e}_{i j}^{(h)}\right)}{\sum_{k=1}^n \exp \left(\mathbf{e}_{i k}^{(h)}\right)} \text {, }\\
z_i^{(h)} & =\sum_{j=1}^n \alpha_{i j}^{(h)}\left(\mathbf{e}_j \mathbf{W}_V^{(h)}\right) \text {, } \; z_i=\operatorname{Concat}\left(z_i^{(1)}, \cdots, z_i^{(H)}\right) \text {, }\\
\tilde{\mathbf{e}}_i & =\operatorname{LayerNorm}\left(\mathbf{e}_i+z_i\right) \text {, }\\
\mathbf{e}'_i & =\operatorname{LayerNorm}\left(\tilde{\mathbf{e}}_i+\operatorname{FC}\left(\operatorname{ReLU}\left(\operatorname{FC}\left(\tilde{\mathbf{e}}_i\right)\right)\right)\right)\text {, }
\end{aligned}
\end{equation}
where each self-attention layer has a fully-connected (FC) layer and a normalization layer (LayerNorm), $H$ is the head number, $1 \leq h \leq H$, and $\mathbf{W}_Q^{(h)}, \mathbf{W}_K^{(h)}, \mathbf{W}_V^{(h)} \in \mathbb{R}^{d_x \times\left(d_x / H\right)}$ are learnable parameter matrices for ``query'', ``key'' and ``value''. Each head of each self-attention layer
provides a learned relation between all the input $\mathbf{e} \in \mathbf{E}$, the relation strength is determined in the attention weights $\alpha_{i j}^{(h)}$.

The second operation Inter-hierarchy Communication, IRC for short.
It is used to realize the message-passing from the child nodes to the parent node.
Given the a node $\mathbf{e}$, its $k$th child node's hidden state and memory cell are denoted as $\mathbf{h}_{\ell}$ and $\mathbf{c}_{\ell}$.
We follow the $N$-ary structure of the Tree-LSTM network \cite{tai2015improved} to perform the hidden state transition among the nodes as follows:
\begin{small}
\begin{gather}
% \begin{equation}
% \begin{aligned}
    \mathbf{i} =\sigma\left(\mathbf{W}^{(\mathbf{i})} \mathbf{e}+\sum_{\ell=1}^{N} \mathbf{U}_{\ell}^{(i)} \mathbf{h}_{\ell}+\mathbf{b}^{(i)}\right)\text {, }\nonumber\\
    \mathbf{f}_{k} =\sigma\left(\mathbf{W}^{(\mathbf{f})} \mathbf{e}+\sum_{\ell=1}^{N} \mathbf{U}_{k \ell}^{(f)} \mathbf{h}_{\ell}+\mathbf{b}^{(f)}\right)\text {, }\nonumber\\
    \mathbf{o} =\sigma\left(\mathbf{W}^{(\mathbf{o})} \mathbf{e}+\sum_{\ell=1}^{N} \mathbf{U}_{\ell}^{(o)} \mathbf{h}_{\ell}+\mathbf{b}^{(o)}\right)\text {, }\\
    \mathbf{u} =\tanh \left(\mathbf{W}^{(\mathbf{u})} \mathbf{e}+\sum_{\ell=1}^{N} \mathbf{U}_{\ell}^{(u)} \mathbf{h}_{\ell}+\mathbf{b}^{(u)}\right)\text {, }\nonumber\\
    \mathbf{c} =\mathbf{i}_{j} \odot \mathbf{u}+\sum_{\ell=1}^{N} \mathbf{f}_{\ell} \odot \mathbf{c}_{\ell}\text {, } \quad  \mathbf{e}'' =\mathbf{o} \odot \tanh \left(\mathbf{c}\right)\text {, }\nonumber
% \end{aligned}
% \end{equation
\end{gather}
\end{small}
% $x_j$ denotes the input of the time step $t$,
where $\sigma$ and $\tanh$ are the sigmoid and hyperbolic tangent activation functions, respectively; $\odot$ is the Hadamard Product; $\mathbf{W}^{(i)}$, $\mathbf{W}^{(f)}$, $\mathbf{W}^{(o)}$, $\mathbf{W}^{(u)}$, $\mathbf{U}_{\ell}^{(i)}$, $\mathbf{U}_{k \ell}^{(f)}$, $\mathbf{U}_{\ell}^{(o)}$, and $\mathbf{U}_{\ell}^{(u)}$ are learnable parameter matrices.

\begin{figure}[t]
    \centering
    \includegraphics[width=0.98\columnwidth]{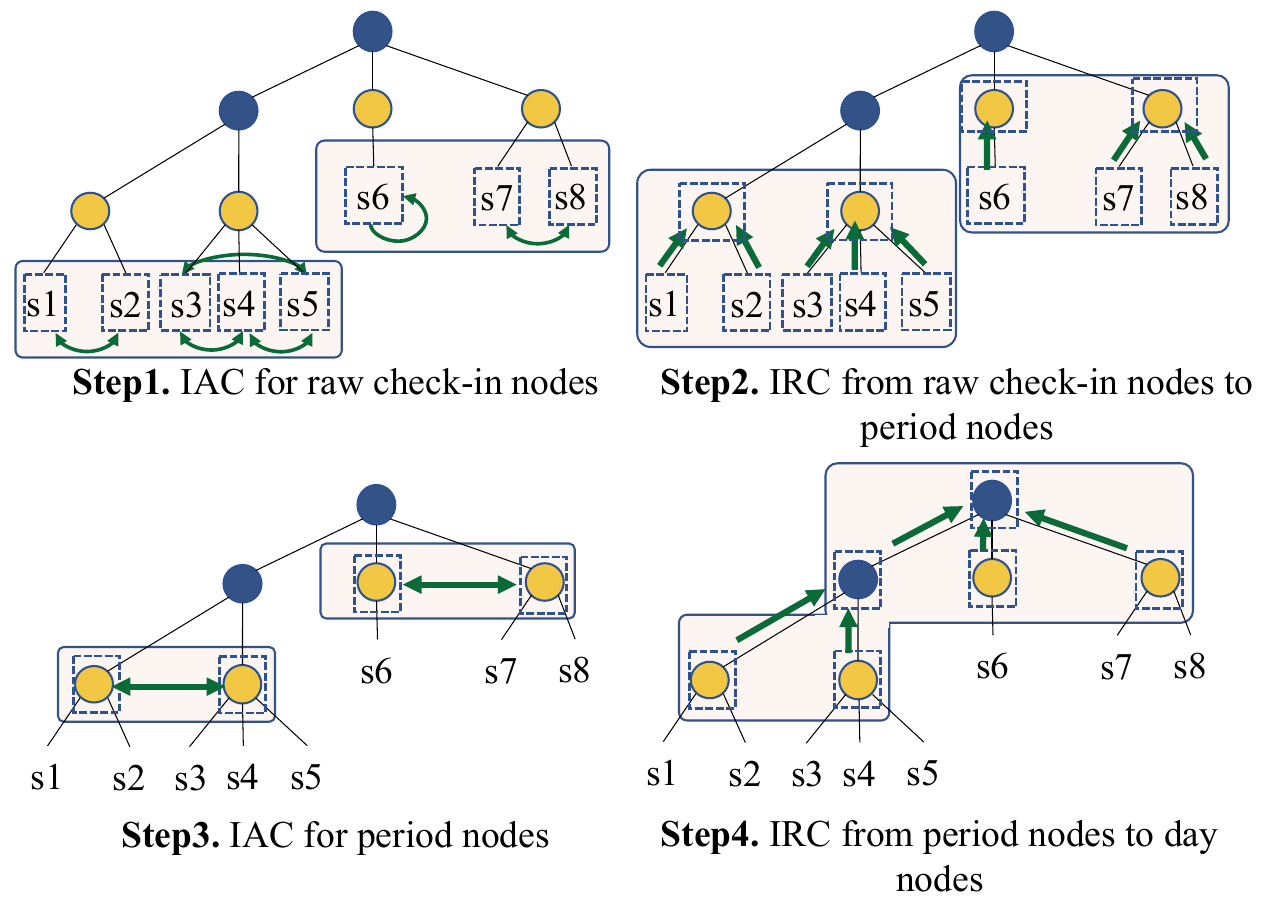}
    \caption{Four-step node interaction.}
    \label{fig_four_steps}
\end{figure}
The node information interaction consists of four steps, as shown in \figurename~\ref{fig_four_steps}.
\textbf{Step 1.} The raw check-in nodes belonging to the same period node exchange information using IAC. This step allow the check-in records can fully capture the features of other check-ins in the same period.
% Using IAC exchanges information among raw check-in nodes belonging to the same period node, so that the check-in records can fully capture the features of other check-ins in the same period.
\textbf{Step 2.} The period nodes process IRC to aggregate the check-in information from their raw check-in nodes.
\textbf{Step 3.} The period nodes belonging to the same-day node exchange information using IAC. This step makes the period preference representations obtain the features from other periods.
% the check-in records can fully capture the features of other check-ins in the same period.
\textbf{Step 4.} The day nodes process IRC to aggregate the period preference from their period nodes.
After four-step node interaction, we obtain the representation of the root node denoted as $\mathbf{e}^{\left(k\right)}$.

\subsection{Multitask Learning}
To push the model towards learning a robust representation, %capability,
we design a multitask training framework to simultaneously predict the next POI, geographical cluster, and category.
\begin{gather}
% \begin{equation}
\hat{\mathbf{y}}_{l}=\mathbf{e}^{\left(k\right)} \mathbf{W}_{l}+\mathbf{b}_{l} \text {, } \quad
\hat{\mathbf{y}}_{g}=\mathbf{e}^{\left(k\right)} \mathbf{W}_{g}+\mathbf{b}_{g} \text {, } \nonumber\\
\hat{\mathbf{y}}_{c}=\mathbf{e}^{\left(k\right)} \mathbf{W}_{c}+\mathbf{b}_{c}\text {, }
% \end{equation}
\end{gather}
where $\mathbf{W}_{l} \in \mathbb{R}^{d\times |\mathcal{L}|} $, $\mathbf{W}_{g} \in \mathbb{R}^{d\times |\mathcal{G}|}$, $\mathbf{W}_{c} \in \mathbb{R}^{d\times |\mathcal{C}|}$, $\mathbf{b}_{l}$, $\mathbf{b}_{g}$, and $\mathbf{b}_{c}$ are learnable parameters for dense layers.

Inspired by \cite{kendall2018multi}, our loss function can be represented as follows:
\begin{equation}
    \mathcal{L}_{\text{final}} = \frac{1}{2 \sigma_{l}^{2}} \mathcal{L}_{l}+\frac{1}{2 \sigma_{g}^{2}} \mathcal{L}_{g}+\frac{1}{2 \sigma_{c}^{2}} \mathcal{L}_{c} +\log \sigma_{l} \sigma_{g} \sigma_{c},
\end{equation}
where $\sigma_{l}, \sigma_{g}$, and $\sigma_{c}$ are learnable parameters, and the last term serves as a regularization term for denoising, $\mathcal{L}_{l}, \mathcal{L}_{g}$, and $\mathcal{L}_{c}$ are cross-entropy loss for next POI, geographical cluster, and category respectively. They are formulated as:
% \begin{equation}
% \begin{gathered}
\begin{gather}
\mathcal{L}_{l} = - \sum_{i=1}^{|\mathcal{L}|} y_{l}^{i} \log \hat{y}_{l}^{i} \text {, } \quad
\mathcal{L}_{g} = -  \sum_{i=1}^{|\mathcal{G}|} y_{g}^{i} \log \hat{y}_{g}^{i} \text {, } \nonumber \\
\mathcal{L}_{c} = -  \sum_{i=1}^{|\mathcal{C}|} y_{c}^{i} \log \hat{y}_{c}^{i} \,,
\end{gather}
where $y_{*}^i$ refers to the $i$-th item of the vector $\mathbf{y}_{*}$.
% \end{gathered}
% \end{equation}

In the recommendation stage, the preference score of the next POI aggregates the prediction scores from the current day node, period node, and last raw check-in node, representing the preferences from the transition relationship, current day, and current period, respectively. The recommendation score is calculated as follows:
\begin{equation}
\hat{\mathbf{y}}_{l}^{\text{rec}}=\eta \hat{\mathbf{y}}_{l}^{\text{day}} + \delta \hat{\mathbf{y}}_{l}^{\text{period}} +  \hat{\mathbf{y}}_{l}^{\text{check-in}} \text {, } 
\end{equation}
where $\eta$ and $\delta$ are used to control the influence of the day and period preference score.

\section{Experiments}
In this section, we conduct a series of experiments to thoroughly evaluate our proposed model MTNet. First, we introduce the datasets and the state-of-the-art baselines. Then, we compare the performance of MTNet with the baselines to show the superiority of our method. Furthermore, we conduct a set of comparative experiments to discuss the impact of time slot factor on our model. We also conduct an ablation experiment to check out the effectiveness of each component in our model. Additionally, we visualize the representations of users and POIs to demonstrate the fine-grained user preferences learned by MTNet.

\subsection{Experimental Setup}
\subsubsection{Datasets}
We conduct experiments using three widely used datasets acquired from two LBSN platforms, namely Foursquare and Gowalla. Specifically, for Foursquare, we use data collected separately in Tokyo \cite{1} and New York City \cite{1} during the period between 12th April 2012 and 16th February 2013. For Gowalla, we utilize data collected in California and Nevada \cite{2} spanning from February 2009 to October 2010. In these three datasets, each row corresponds to a check-in record, encompassing user ID, POI ID, POI category, check-in time, GPS coordinates.

Following previous study \cite{10}, we exclude inactive users who have fewer than 10 check-in records. We also eliminate unpopular POIs that are visited less than 10 times by the remaining users. Considering that a substantial time interval between two records weakens the temporal dependency between them, we divide the complete check-in sequences of users into trajectories by 24 hours. Additionally, with the requirement to convert trajectories into a Mobility Tree, %later on, 
we discard all trajectories with a length of less than 2. The statistical details of the processed datasets are presented in Table \ref{tab:datasets}.

\begin{table}[htbp]
\small
\centering

\begin{tabular}{ccccc}
\toprule 
\textbf{dataset} & \textbf{user} & \textbf{POI} & \textbf{check-in} & \textbf{trajectory} \\
\midrule 
\text{NYC} & 1,075 & 5,099 & 104,074 & 14,160 \\
\text{TKY} & 2,281 & 7,844 & 361,430 & 44,692 \\
\text{CA} & 4,318 & 9,923 & 250,780 & 32,920 \\
\bottomrule
\end{tabular}

\caption{Statistics of the three datasets.}
\label{tab:datasets}

\end{table}

Subsequently, We partition the datasets into training, validation, and test sets in chronological order. The training set, consisting of the initial 80\% of check-ins, is used to train the model. Subsequently, the middle 10\% of check-ins form the validation set, which is utilized for selecting the best-performing model. Finally, we evaluate the model on the test set that consists of the last 10\% of check-ins.

\begin{table*}[t]
\small
\centering

\setlength{\tabcolsep}{5pt}
\begin{tabular}{ccccc|cccc|cccc}
  \toprule
  \multirow{2}{*}{\text{Models}} 
  & \multicolumn{4}{c}{\textbf{NYC}} 
  & \multicolumn{4}{c}{\textbf{TKY}} 
  & \multicolumn{4}{c}{\textbf{CA}} \\
  & \text{$Acc@1$} & \text{$Acc@5$} & \text{$Acc@10$} & \text{$MRR$} 
  & \text{$Acc@1$} & \text{$Acc@5$} & \text{$Acc@10$} & \text{$MRR$} 
  & \text{$Acc@1$} & \text{$Acc@5$} & \text{$Acc@10$} & \text{$MRR$} \\
  \midrule
  \text{FPMC} 
  & 0.1003 & 0.2126 & 0.2970 & 0.1701 
  & 0.0814 & 0.2045 & 0.2746 & 0.1344
  & 0.0383 & 0.0702 & 0.1159 & 0.0911 \\
  \text{PRME} 
  & 0.1159 & 0.2236 & 0.3105 & 0.1712 
  & 0.1052 & 0.2728 & 0.2944 & 0.1786 
  & 0.0521 & 0.1034 & 0.1425 & 0.1002 \\
  \text{LSTM} 
  & 0.1305 & 0.2719 & 0.3283 & 0.1857 
  & 0.1335 & 0.2728 & 0.3277 & 0.1834 
  & 0.0665 & 0.1306 & 0.1784 & 0.1201 \\
  \text{ST-RNN} 
  & 0.1483 & 0.2923 & 0.3622 & 0.2198 
  & 0.1409 & 0.3022 & 0.3577 & 0.2212 
  & 0.0799 & 0.1423 & 0.1940 & 0.1429 \\
  \text{STGN} 
  & 0.1716 & 0.3381 & 0.4122 & 0.2598 
  & 0.1689 & 0.3391 & 0.3848 & 0.2422 
  & 0.0810 & 0.1842 & 0.2579 & 0.1675 \\
  \text{STGCN} 
  & 0.1799 & 0.3425 & 0.4279 & 0.2788 
  & 0.1716 & 0.3453 & 0.3927 & 0.2504 
  & 0.0961 & 0.2097 & 0.2613 & 0.1712 \\
  \text{PLSPL} 
  & 0.1917 & 0.3678 & 0.4523 & 0.2806 
  & 0.1889 & 0.3523 & 0.4150 & 0.2542 
  & 0.1072 & 0.2278 & 0.2995 & 0.1847 \\
  \text{CFPRec} 
  & 0.1692 & 0.3867 & 0.4894 & 0.2680 
  & 0.2052 & 0.4028 & 0.4769 & 0.2963 
  & 0.0473 & 0.1420 & 0.1874 & 0.0911 \\
  \text{STAN} 
  & 0.2231 & 0.4582 & 0.5734 & 0.3253 
  & 0.1963 & 0.3798 & 0.4464 & 0.2852 
  & 0.1104 & 0.2348 & 0.3018 & 0.1869 \\
  \text{GETNext} 
  & \underline{0.2435} & \underline{0.5089} & \underline{0.6143} & \underline{0.3621} 
  & \underline{0.2254} & \underline{0.4417} & \underline{0.5287} & \underline{0.3262} 
  & \underline{0.1357} & \underline{0.2852} & \underline{0.3590} & \underline{0.2103} \\
  \midrule
  \rowcolor{gray!25}
  \text{MTNet}
  & \textbf{0.2620} & \textbf{0.5381} & \textbf{0.6321} & \textbf{0.3855} 
  & \textbf{0.2575} & \textbf{0.4977} & \textbf{0.5848} & \textbf{0.3659} 
  & \textbf{0.1453} & \textbf{0.3419} & \textbf{0.4163} & \textbf{0.2367} \\
  \rowcolor{gray!25}
  \textit{Impro} 
  & 7.60\% & 5.74\% & 2.90\% & 6.46\% 
  & 14.24\% & 12.68\% & 10.61\% & 12.17\% 
  & 9.45\% & 19.88\% & 15.96\% & 12.55\% \\
  \bottomrule
\end{tabular}

\caption{Performance of the model on the NYC, TKY and CA datasets compared based on the Accuracy ($Acc$) and Mean Reciprocal Rank ($MRR$) metrics. We present the results in ascending order based on the model's performance, highlighting the best results in bold, and underlining the second best results.}
\label{tab:main_comparison}
\end{table*}

\subsubsection{Baseline Models}
We compare MTNet with ten state-of-the-art methods: 1) \textbf{FPMC} \cite{4}: integrates Matrix Factorization and Markov Chains; 2) \textbf{PRME} \cite{5}: introduces a pair-wise metric embedding method; 3) \textbf{LSTM} \cite{3}: is a modified RNN architecture; 4) \textbf{ST-RNN} \cite{6}: extends RNN by considering time and distance transition matrices; 5) \textbf{STGN} \cite{7}: enhances LSTM by adding spatiotemporal gates; 6) \textbf{STGCN} \cite{7}: enhances STGN by using coupled input and forget gates; 7) \textbf{PLSPL} \cite{8}: employs attention mechanism and utilizes LSTM; 8) \textbf{CFPRec} \cite{zhang2022next}: models users' past, present and future preferences; 9) \textbf{STAN} \cite{9}: is a spatiotemporal bi-attention model; 10) \textbf{GETNext} \cite{10}: introduces a trajectory flow map to capture spatial transition information.

\subsubsection{Evaluation Metrics}
We utilize two commonly employed evaluation metrics, namely average Accuracy@$K$ ($Acc@K$) and Mean Reciprocal Rank ($MRR$), to evaluate the effectiveness of the recommendation model. Both $Acc@K$ and $MRR$ serve as positive indicators.

\subsubsection{Experiment Settings}
We develop MTNet\footnote{\url{https://github.com/Skyyyy0920/MTNet}} based on PyTorch and conduct experiments on hardware with AMD Ryzen 7 4800H CPU and NVIDIA GeForce RTX 2060 GPU. We keep the hyper-parameters consistent on NYC, TKY and CA. We set the number of time slots to 12 for TKY and CA, and 4 for NYC, according to the performance on the validation set. The user and POI embedding dimensions are both set to 128, while the category and geography embedding dimensions are set to 32. The hidden size for Tree-LSTM module is 512. We employ the Adam \cite{kingma2014adam} optimizer with an initial learning rate of $1 \times 10^{-3}$ and a weight decay rate of $1 \times 10^{-4}$. We set the influence of day node $\eta=1$ and period node $\delta=1$. We generate 60 clusters for geographical information representation. Moreover, we utilize a step-by-step learning rate scheduler with a step size of 6 and a decay factor of 0.9. For the Transformer component, we incorporate 2 transformer layers, each consisting of 2 attention heads and a dimension of 1024. Additionally, we randomly drop the embeddings and parameters with a dropout rate of 0.4 and 0.6 respectively. Finally, we run each model for a total of 50 epochs with a batch size of 1024.

\subsection{Results and Analysis}

We compare the performance of our proposed model and the baselines on the three datasets. We take $Acc@1$, $Acc@5$, $Acc@10$ and $MRR$ as the evaluation metrics. For all models, we use the same data preprocessing method following GETNext \cite{10}. The experimental results presentd in Table \ref{tab:main_comparison} show that our proposed model MTNet outperformes all other state-of-the-art baselines across all datasets in terms of all evaluation metrics. 

We find that all models perform the best on the NYC dataset, followed by TKY, while the performance on the CA dataset was the worst. The primary reason for this could be that the NYC and TKY datasets collect check-in records of a limited number of users in a small geographical area. These two datasets have fewer POIs and a smaller user population. In contrast, the CA dataset is collected across the larger regions of California and Nevada, with the most users (4318, four times that of NYC) and POIs (9923, nearly twice as many as in NYC). Therefore, CA presents a challenge for all recommendation models. The results show that MTNet performs exceptionally well on the CA dataset. It averages a superior performance over the second-ranked model, GETNext, by 15.26\% in terms of $Acc$, and it exhibits a notable improvement of 12.55\% in terms of $MRR$. The improvement once again demonstrates that MTNet effectively learns long-term and short-term preferences of the data. This is primarily due to the average trajectory length in the training set of CA being 8.68, whereas the value is 7.55 for NYC and 8.22 for TKY. This enhances the advantage of MTNet over other models in segregating user preferences across different time slots from the users' check-in trajectories.

It can be also observed that GETNext achieves the second best performance on the three datasets. 
GETNext designs a trajectory flow map to effectively capture common movement patterns of users, which tackle challenges related to inactive users and short trajectories well, while it is not able to learn fine-grained user preferences at different time slots. Besides, STAN also performs well on the three datasets using a bi-attention model to learn the spatio-temporal correlations of user's trajectory.

\subsection{Impact of Time Slot Factor}
In the implementation, we convert timestamps to integers ranging from 1 to 24 for check-in time embedding learning. Then, the integer times are assigned to their corresponding time slots. 
To investigate the influence of the time slot factor on MTNet, we conduct experiments with various numbers of period nodes.

\begin{figure}[htbp]
    \centering
    \includegraphics[width=0.98\columnwidth]{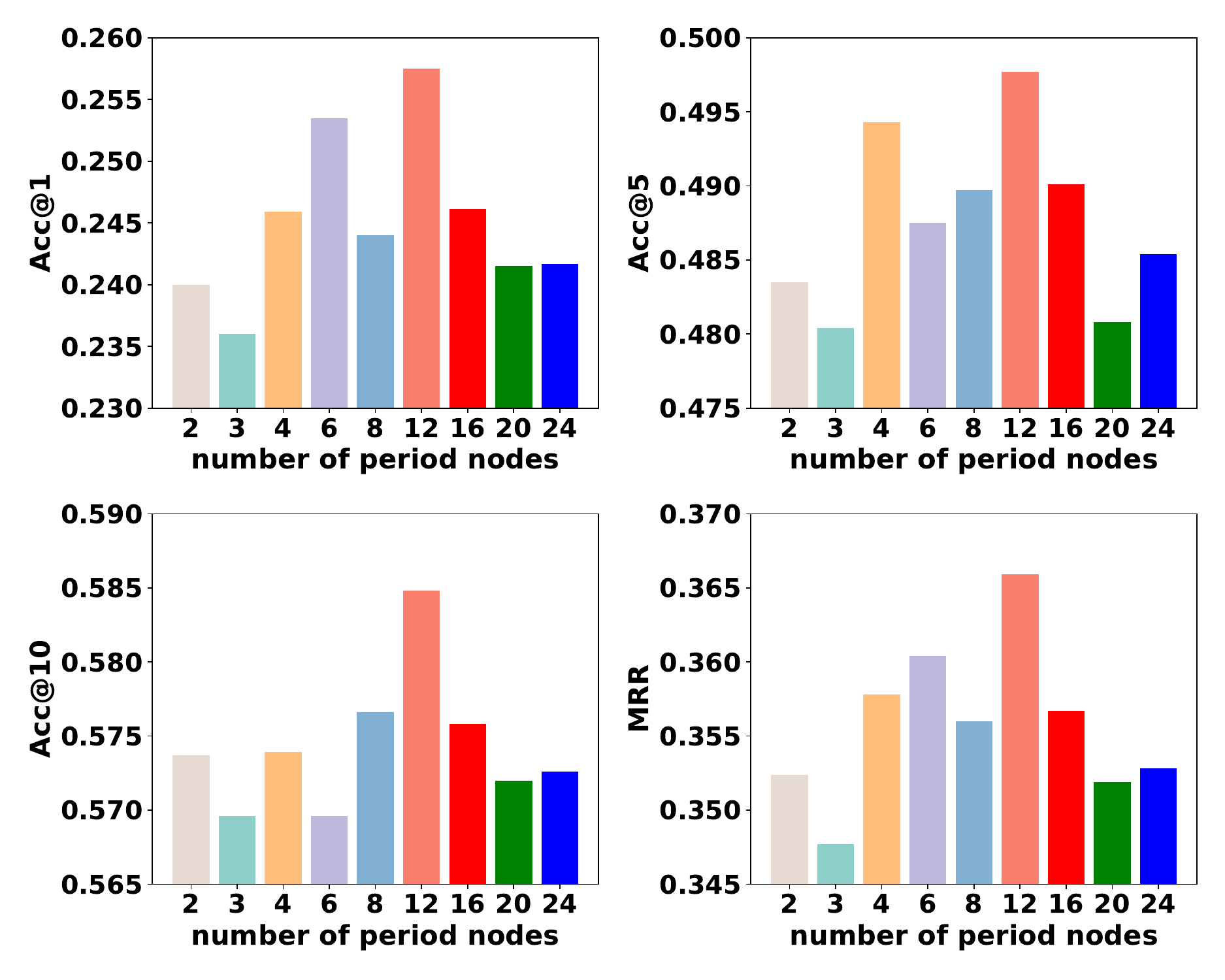}
    \caption{Performance of MTNet with different period node numbers of 2, 3, 4, 6, 8, 12, 16, 20 and 24 on TKY.}
    \label{fig:time_slot_impact_on_TKY}
\end{figure}

It can be observed from Figure~\ref{fig:time_slot_impact_on_TKY} that the performance with 12 period nodes outperforms other values across all metrics, where each period is 2-hour length. From empirical analysis on TKY dataset, we figure out that the visitation time window (the interval between the earliest and latest check-in times) for users to check-in the same POI is around 2 hours. The observation of the dataset aligns with the experimental results, indicating that the number of time slots used to construct the Mobility Tree should correspond to the density of user check-in timestamps. Moreover, when the time slot is set to 2, 3, 20, or 24, we observe a consistently poor performance of the model across all metrics. This suggests that if we divide the day into only 2 or 3 time periods (i.e., daytime and nighttime) or if we excessively fine-grain the time intervals (such as every hour), the model's ability to learn user preferences at specific times diminishes.

\subsection{Ablation Study}
In this section, we conduct an ablation study to evaluate the contributions of each component of MTNet. We perform a total of eight experiments on all three datasets: 1) Full model; 2) Model without multi-task learning mechanism, which implies that the model's loss function is simply the direct summation of the losses generated from predictions for POI, category, and geography; 3) Model without geography prediction; 4) Model without category prediction; 5) Model without multi-objective prediction task (Remove both category and geography prediction); 6) Model without IAC; 7) Model without IRC; 8) Model without current period node and last check-in node prediction. The experimental results on TKY is documented in Table \ref{tab:ablation_TKY}.

It can be observed that the full model significantly surpasses other variants. Furthermore, we can observe that IAC and IRC have a more pronounced impact on the overall model performance. The model's performance drops by 8.63\% and 7.14\% after removing the IAC and IRC modules, respectively, highlighting the substantial influence of IAC and IRC on the model's ability to learn user trajectory preferences across different time slots. When the model does not employ multi-task learning, we can observe that the average performance of the model decreases by 4.52\%. This indicates that the multi-task learning mechanism has a significant impact on the model's training, as it helps the model learn more crucial information of check-ins. In addition, when we remove the auxiliary predictions of the current period node and last raw check-in node, we find a certain degree of performance decline as well. This shows that, during recommendation stage, the model not only relies on the user's historical preferences but also heavily considers the most recent trajectory for predicting the next POI.

\begin{table}[htbp]
\small
\centering

\begin{tabular}{ccccc}
  \toprule
  \text{Variants}
  & \text{$Acc@1$} & \text{$Acc@5$} & \text{$Acc@10$} & \text{$MRR$} \\
  \midrule
  \rowcolor{gray!25}
  \text{Full Model} &
  \textbf{0.2575} & \textbf{0.4977} & \textbf{0.5848} & \textbf{0.3659} \\
  \midrule
  \text{$w/o$ multitask} & 
  0.2410 & 0.4770 & 0.5631 & 0.3499 \\
  \text{$w/o$ coo} & 
  0.2353 & 0.4766 & 0.5595 & 0.3453 \\ 
  \text{$w/o$ cat} & 
  0.2398 & 0.4799 & 0.5659 & 0.3484 \\
  \text{$w/o$ coo\&cat} & 
  0.2377 & 0.4791 & 0.5669 & 0.3474 \\ 
  \text{$w/o$ IAC} & 
  0.2353 & 0.4757 & 0.5631 & 0.3469 \\
  \text{$w/o$ IRC} & 
  0.2391 & 0.4713 & 0.5504 & 0.3453 \\ 
  \text{$w/o$ node} & 
  0.2478 & 0.4806 & 0.5686 & 0.3567 \\
  \bottomrule
\end{tabular}

\caption{Ablation studies on TKY. The best results are highlighted in bold.}
\label{tab:ablation_TKY}
\end{table}

\subsection{Visualization}
\begin{figure}[t]
    \centering
    \begin{subfigure}[b]{0.48\linewidth}
        \centering
        \includegraphics[width=\linewidth]{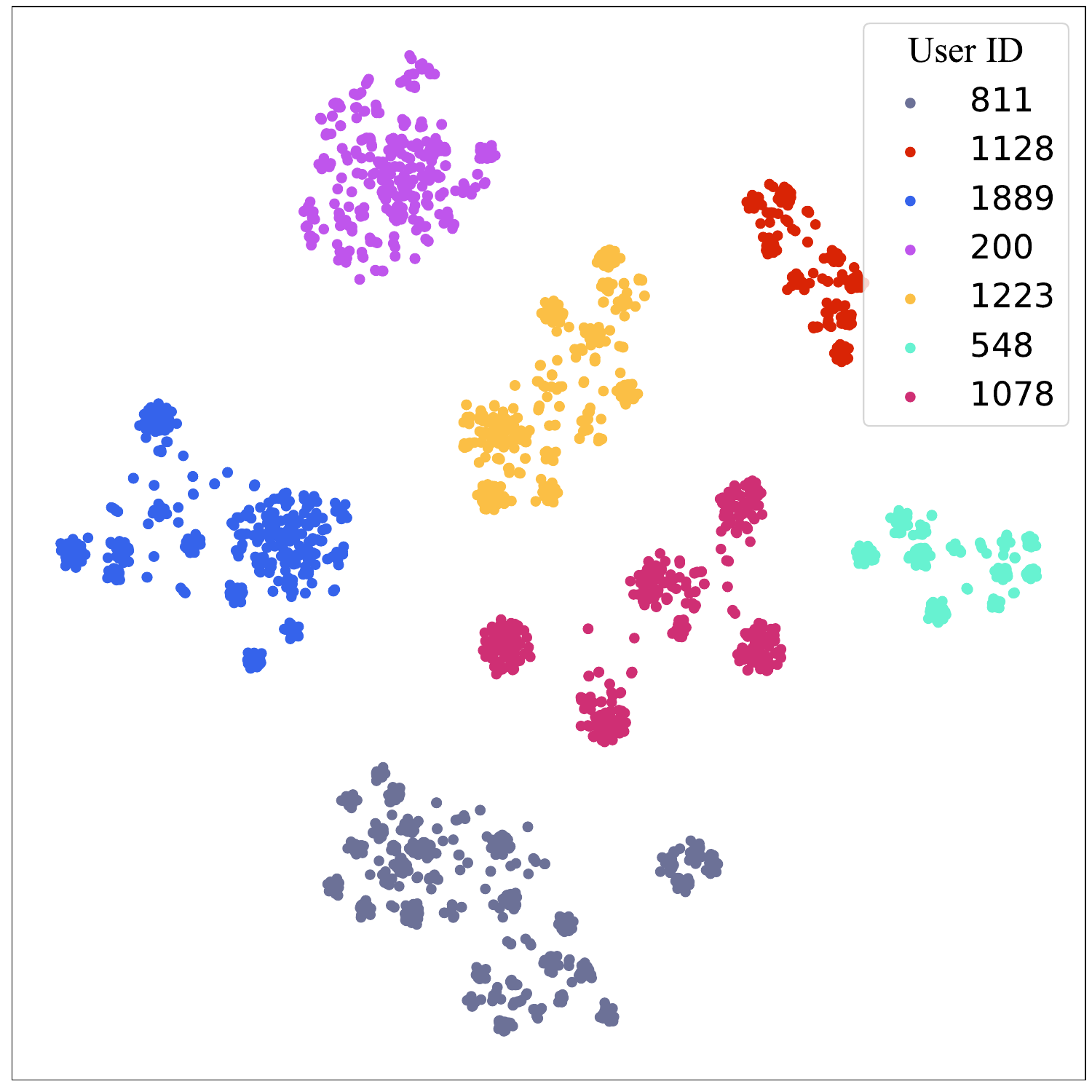}
        \caption{}
        \label{fig:7users}
    \end{subfigure}
    % \hspace{5mm}
    \begin{subfigure}[b]{0.48\linewidth}
        \centering
        \includegraphics[width=\linewidth]{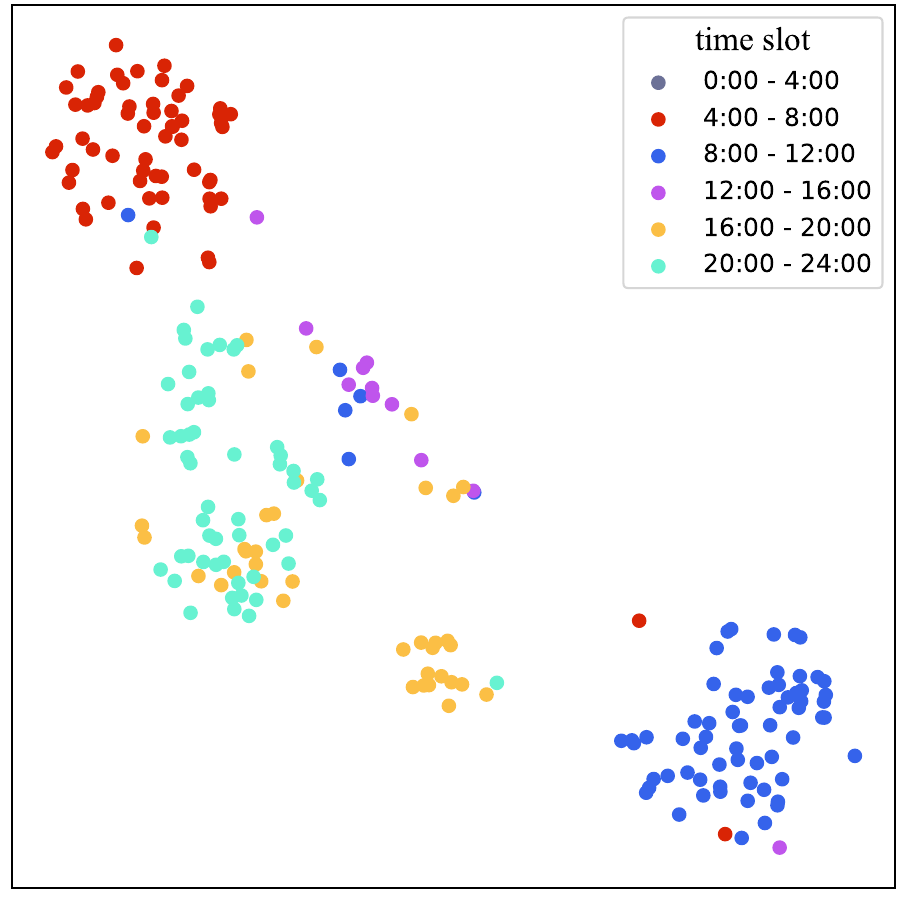}
        \caption{}
        \label{fig:user_h}
    \end{subfigure}
    
    \caption{(a) Visualization of different users within a unified time period (20:00-22:00) on TKY. (b) Visualization of clustering results for a user's trajectories at different periods.}
    \label{fig:visualization}
\end{figure}

To enhance the understanding of MTNet, we conduct visualizations of different users' representations during the same time period, as well as visualizations of the representations of the same user's trajectories across different time periods. 
We conduct an experiment to visualize trajectory embeddings for all users. While we can observe effective differentiation among users, the sheer volume of users (over 2000) makes it hard to discern individual patterns. Therefore, we intentionally selected users with a substantial trajectory count (over 150) to ensure a clear presentation, as depicted in Figure \ref{fig:7users}. We can observe that during the same time period of 20:00-22:00, the representations of different users are effectively distinguished. This indicates that MTNet has indeed learned the distinct preferences of various users during the same time period. In Figure \ref{fig:user_h}, as shown, when a user is in different time periods, their trajectory representations also exhibit distinct variations. This demonstrates that the model can learn the unique preferences of users during different time slots.

\section{Conclusion}
This paper introduces MTNet, which leverages a novel user check-in description structure of Mobility Tree for the next POI recommendation.
The Mobility Tree's distinctive attribute lies in its multi-granularity time slot nodes, specially designed to encapsulate users' diverse preferences across different periods.
We propose a four-step node interaction operation, which facilitates the comprehensive propagation and aggregation of check-in features, traversing from the leaf nodes to the root node.
In pursuit of a more robust representation, MTNet adopts a multitasking training strategy that involves the simultaneous prediction of the next POI with the contextual information, thereby improving the recommendation performance.
Our experiments on three real-world LBSN datasets suggested that MTNet outperforms ten state-of-the-art methods for the next POI recommendation.
%
%Moreover, we use the visualization method to display users' diverse time slot preferences.
%
For future work, we will expand the tree structure with heterogeneous nodes to facilitate more spatial-temporal context interaction for time slot preference exploration.

\section{Acknowledgments}
This work was supported by the National Key R\&D Program of China (2022YFB3103202), the National Natural Science Foundation of China (No. U1936206, 62002178, 62272250), and the Natural Science Foundation of Tianjin, China (No. 22JCJQJC00150).

\bibliography{aaai24}

\begin{thebibliography}{26}
\providecommand{\natexlab}[1]{#1}

\bibitem[{Cho, Myers, and Leskovec(2011)}]{2}
Cho, E.; Myers, S.~A.; and Leskovec, J. 2011.
\newblock Friendship and Mobility: User Movement in Location-Based Social Networks.
\newblock In \emph{Proceedings of the 17th ACM SIGKDD International Conference on Knowledge Discovery and Data Mining}, KDD '11, 1082–1090. New York, NY, USA: Association for Computing Machinery.
\newblock ISBN 9781450308137.

\bibitem[{Feng et~al.(2018)Feng, Li, Zhang, Sun, Meng, Guo, and Jin}]{feng2018deepmove}
Feng, J.; Li, Y.; Zhang, C.; Sun, F.; Meng, F.; Guo, A.; and Jin, D. 2018.
\newblock Deepmove: Predicting human mobility with attentional recurrent networks.
\newblock In \emph{Proceedings of the 2018 world wide web conference}, 1459--1468.

\bibitem[{Feng et~al.(2015)Feng, Li, Zeng, Cong, and Chee}]{5}
Feng, S.; Li, X.; Zeng, Y.; Cong, G.; and Chee, Y. 2015.
\newblock Personalized ranking metric embedding for next new POI recommendation.
\newblock In \emph{IJCAI'15 Proceedings of the 24th International Conference on Artificial Intelligence}, 2069--2075. ACM.

\bibitem[{Guo et~al.(2020)Guo, Sun, Zhang, and Theng}]{guo2020attentional}
Guo, Q.; Sun, Z.; Zhang, J.; and Theng, Y.-L. 2020.
\newblock An attentional recurrent neural network for personalized next location recommendation.
\newblock In \emph{Proceedings of the AAAI Conference on artificial intelligence}, volume~34, 83--90.

\bibitem[{Hochreiter and Schmidhuber(1997)}]{3}
Hochreiter, S.; and Schmidhuber, J. 1997.
\newblock Long Short-Term Memory.
\newblock \emph{Neural Computation}, 9(8): 1735--1780.

\bibitem[{Huang et~al.(2019)Huang, Ma, Wang, and Liu}]{huang2019attention}
Huang, L.; Ma, Y.; Wang, S.; and Liu, Y. 2019.
\newblock An attention-based spatiotemporal lstm network for next poi recommendation.
\newblock \emph{IEEE Transactions on Services Computing}, 14(6): 1585--1597.

\bibitem[{Kendall, Gal, and Cipolla(2018)}]{kendall2018multi}
Kendall, A.; Gal, Y.; and Cipolla, R. 2018.
\newblock Multi-task learning using uncertainty to weigh losses for scene geometry and semantics.
\newblock In \emph{Proceedings of the IEEE conference on computer vision and pattern recognition}, 7482--7491.

\bibitem[{Kingma and Ba(2014)}]{kingma2014adam}
Kingma, D.~P.; and Ba, J. 2014.
\newblock Adam: A method for stochastic optimization.
\newblock \emph{arXiv preprint arXiv:1412.6980}.

\bibitem[{Lian et~al.(2020)Lian, Wu, Ge, Xie, and Chen}]{lian2020geography}
Lian, D.; Wu, Y.; Ge, Y.; Xie, X.; and Chen, E. 2020.
\newblock Geography-aware sequential location recommendation.
\newblock In \emph{Proceedings of the 26th ACM SIGKDD international conference on knowledge discovery \& data mining}, 2009--2019.

\bibitem[{Lim et~al.(2022)Lim, Hooi, Ng, Goh, Weng, and Tan}]{lim2022hierarchical}
Lim, N.; Hooi, B.; Ng, S.-K.; Goh, Y.~L.; Weng, R.; and Tan, R. 2022.
\newblock Hierarchical multi-task graph recurrent network for next poi recommendation.
\newblock In \emph{Proceedings of the 45th international ACM SIGIR conference on Research and development in Information Retrieval}.

\bibitem[{Liu et~al.(2016)Liu, Wu, Wang, and Tan}]{6}
Liu, Q.; Wu, S.; Wang, L.; and Tan, T. 2016.
\newblock Predicting the Next Location: A Recurrent Model with Spatial and Temporal Contexts.
\newblock \emph{Proceedings of the AAAI Conference on Artificial Intelligence}, 30(1).

\bibitem[{Liu et~al.(2017)Liu, Pham, Cong, and Yuan}]{liu2017experimental}
Liu, Y.; Pham, T.-A.~N.; Cong, G.; and Yuan, Q. 2017.
\newblock An experimental evaluation of point-of-interest recommendation in location-based social networks.
\newblock \emph{Proceedings of the VLDB Endowment}, 10(10): 1010--1021.

\bibitem[{Luo, Liu, and Liu(2021)}]{9}
Luo, Y.; Liu, Q.; and Liu, Z. 2021.
\newblock STAN: Spatio-Temporal Attention Network for Next Location Recommendation.
\newblock In \emph{Proceedings of the Web Conference 2021}, WWW '21, 2177–2185. New York, NY, USA: Association for Computing Machinery.
\newblock ISBN 9781450383127.

\bibitem[{Manotumruksa, Macdonald, and Ounis(2018)}]{manotumruksa2018contextual}
Manotumruksa, J.; Macdonald, C.; and Ounis, I. 2018.
\newblock A contextual attention recurrent architecture for context-aware venue recommendation.
\newblock In \emph{The 41st international ACM SIGIR conference on research \& development in information retrieval}, 555--564.

\bibitem[{Rao et~al.(2022)Rao, Chen, Liu, Shang, Yao, and Han}]{rao2022graph}
Rao, X.; Chen, L.; Liu, Y.; Shang, S.; Yao, B.; and Han, P. 2022.
\newblock Graph-flashback network for next location recommendation.
\newblock In \emph{Proceedings of the 28th ACM SIGKDD Conference on Knowledge Discovery and Data Mining}, 1463--1471.

\bibitem[{Rendle, Freudenthaler, and Schmidt-Thieme(2010)}]{4}
Rendle, S.; Freudenthaler, C.; and Schmidt-Thieme, L. 2010.
\newblock Factorizing Personalized Markov Chains for Next-Basket Recommendation.
\newblock In \emph{Proceedings of the 19th International Conference on World Wide Web}, WWW '10, 811–820. New York, NY, USA: Association for Computing Machinery.
\newblock ISBN 9781605587998.

\bibitem[{S{\'a}nchez and Bellog{\'\i}n(2022)}]{sanchez2022point}
S{\'a}nchez, P.; and Bellog{\'\i}n, A. 2022.
\newblock Point-of-interest recommender systems based on location-based social networks: a survey from an experimental perspective.
\newblock \emph{ACM Computing Surveys (CSUR)}, 54(11s): 1--37.

\bibitem[{Sun et~al.(2020)Sun, Qian, Chen, Liang, Nguyen, and Yin}]{sun2020go}
Sun, K.; Qian, T.; Chen, T.; Liang, Y.; Nguyen, Q. V.~H.; and Yin, H. 2020.
\newblock Where to go next: Modeling long-and short-term user preferences for point-of-interest recommendation.
\newblock In \emph{Proceedings of the AAAI Conference on Artificial Intelligence}, volume~34, 214--221.

\bibitem[{Tai, Socher, and Manning(2015)}]{tai2015improved}
Tai, K.~S.; Socher, R.; and Manning, C.~D. 2015.
\newblock Improved semantic representations from tree-structured long short-term memory networks.
\newblock \emph{arXiv preprint arXiv:1503.00075}.

\bibitem[{Vaswani et~al.(2017)Vaswani, Shazeer, Parmar, Uszkoreit, Jones, Gomez, Kaiser, and Polosukhin}]{vaswani2017attention}
Vaswani, A.; Shazeer, N.; Parmar, N.; Uszkoreit, J.; Jones, L.; Gomez, A.~N.; Kaiser, {\L}.; and Polosukhin, I. 2017.
\newblock Attention is all you need.
\newblock \emph{Advances in neural information processing systems}, 30.

\bibitem[{Wang et~al.(2021)Wang, Wang, Xiang, Yu, Deng, and Xu}]{wang2021attentive}
Wang, D.; Wang, X.; Xiang, Z.; Yu, D.; Deng, S.; and Xu, G. 2021.
\newblock Attentive sequential model based on graph neural network for next poi recommendation.
\newblock \emph{World Wide Web}, 24(6): 2161--2184.

\bibitem[{Wu et~al.(2022)Wu, Li, Zhao, and Qian}]{8}
Wu, Y.; Li, K.; Zhao, G.; and Qian, X. 2022.
\newblock Personalized Long- and Short-term Preference Learning for Next POI Recommendation.
\newblock \emph{IEEE Transactions on Knowledge and Data Engineering}, 34(4): 1944--1957.

\bibitem[{Yang et~al.(2015)Yang, Zhang, Zheng, and Yu}]{1}
Yang, D.; Zhang, D.; Zheng, V.~W.; and Yu, Z. 2015.
\newblock Modeling User Activity Preference by Leveraging User Spatial Temporal Characteristics in LBSNs.
\newblock \emph{IEEE Transactions on Systems, Man, and Cybernetics: Systems}, 45(1): 129--142.

\bibitem[{Yang, Liu, and Zhao(2022)}]{10}
Yang, S.; Liu, J.; and Zhao, K. 2022.
\newblock GETNext: Trajectory Flow Map Enhanced Transformer for Next POI Recommendation.
\newblock In \emph{Proceedings of the 45th International ACM SIGIR Conference on Research and Development in Information Retrieval}, SIGIR '22, 1144–1153. New York, NY, USA: Association for Computing Machinery.
\newblock ISBN 9781450387323.

\bibitem[{Zhang et~al.(2022)Zhang, Sun, Wu, Zhang, Ong, and Qu}]{zhang2022next}
Zhang, L.; Sun, Z.; Wu, Z.; Zhang, J.; Ong, Y.~S.; and Qu, X. 2022.
\newblock Next point-of-interest recommendation with inferring multi-step future preferences.
\newblock In \emph{IJCAI}, 3751--3757.

\bibitem[{Zhao et~al.(2022)Zhao, Luo, Liu, Xu, Li, Zhuang, Sheng, and Zhou}]{7}
Zhao, P.; Luo, A.; Liu, Y.; Xu, J.; Li, Z.; Zhuang, F.; Sheng, V.~S.; and Zhou, X. 2022.
\newblock Where to Go Next: A Spatio-Temporal Gated Network for Next POI Recommendation.
\newblock \emph{IEEE Transactions on Knowledge and Data Engineering}, 34(5): 2512--2524.

\end{thebibliography}

\end{document}